\documentclass{article}

\usepackage{amsmath}
\usepackage{algorithm}
\usepackage{algpseudocode}
\usepackage{amsfonts}
\usepackage{amssymb}
\usepackage[font=small,skip=0pt]{caption}
\usepackage{multirow}
\usepackage{authblk}

\usepackage[english]{babel}
\usepackage{amssymb}

\usepackage[dvipsnames]{xcolor}
\usepackage{soul}

\usepackage[letterpaper,top=2cm,bottom=2cm,left=3cm,right=3cm,marginparwidth=1.75cm]{geometry}

\usepackage{amsmath}
\usepackage{graphicx}
\usepackage[colorlinks=true, allcolors=blue]{hyperref}

\title{Synthetic Aperture Radar Image Segmentation with Quantum Annealing}
\author[1,2]{Timothé Presles}
\author[1]{Cyrille Enderli}
\author[2]{Gilles Burel}
\author[2]{El Houssaïn Baghious}
\affil[1]{Thales Defense Mission Systems, 2 av. Gay Lussac, 78851 Elancourt, France}
\affil[2]{Univ. Brest, Lab-STICC, CNRS, UMR 6285, CyberIoT Chair, 6 av. Le Gorgeu, 29200 Brest, France}

\begin{document}

\maketitle
 
\begin{abstract}
In image processing, image segmentation is the process of partitioning a digital image into multiple image segment. Among state-of-the-art methods, Markov Random Fields (MRF) can be used to model dependencies between pixels, and achieve a segmentation by minimizing an associated cost function. Currently, finding the optimal set of segments for a given image modeled as a MRF appears to be NP-hard. In this paper, we aim to take advantage of the exponential scalability of quantum computing to speed up the segmentation of Synthetic Aperture Radar (SAR) images. For that purpose, we propose a hybrid quantum annealing classical optimization Expectation Maximization algorithm to obtain optimal sets of segments. After proposing suitable formulations, we discuss the performances and the scalability of our approach on the D-Wave quantum computer. We also propose a short study of optimal computation parameters to enlighten the limits and potential of the adiabatic quantum computation to solve large instances of combinatorial optimization problems.
\end{abstract}

\section{Introduction}

Synthetic Aperture Radar (SAR) is an imaging technique that utilizes radar signals to create high-resolution, three-dimensional reconstructions of terrain and objects by exploiting the motion of the radar antenna and signal processing algorithms \cite{Intro_SAR}. Images obtained with SARs cover dozens of kilometers with a resolution between 15 and 30cm, resulting in images of billions of pixels. Due to the low interpretability and noisiness of SAR images, radar operators are using Automatic Target Detection (ATD) algorithms. The goal of these algorithms is to associate a group of pixels to an object of interest, as a vehicle or a radar station, by segmenting the image in different regions. 
In real time area surveillance, due to the size and the high number of images coming from multiple sensors, current computing capabilities and the restrained number of qualified radar operators greatly limits the number images that can be processed.\\
Image segmentation is the process to divide an image in multiple classes, called segments or regions, in order to extract information about the image. Segmentation results takes the form of a set of labelled pixels, where each label corresponds to a class. The situation where the true label of each pixel is not known is sometimes called unsupervised image segmentation. In the literature, unsupervised image segmentation can be done with region smoothing \cite{region_smoothing}, region growing \cite{region_growing}, clustering \cite{clustering, fuzzy_clustering}, watershed segmentation \cite{watershed}, graph-based methods \cite{graph_based}. Due to the exponentially increasing number of possible labellings, solving large instances of image segmentation problems often result in sub-optimal sets of labels. \\
Quantum computing is an emerging technology which exploits the laws of quantum mechanics in order to perform logical operations. Instead of classical bits, quantum computers operate on qubits, which are in a superposition of two states. Among various applications, solving optimisation problems with quantum devices promises an exponential speedup over classical approaches \cite{los_alamos}, especially for large instances of NP-Hard problems.  There are currently two main approaches in the design of quantum computers: Circuit-oriented quantum computers and quantum annealers. Circuit oriented quantum computers have a sequential approach of quantum computation, using gates to perform operations on single or multiple qubits. Quantum annealers have a simultaneous approach of quantum computation, making all the qubits involved in the computation converge from an initial state to a final state. In this work, we aim to propose an approach to speed up the processing of a high number of high resolution SAR images in order to tackle the current limitation regarding the analysing capability. Beyond this latter problem, our approach could also contribute for the training of Automatic Target Recognition (ATR) classifiers \cite{ATR_SAR}, by providing large datasets of segmented SAR images obtained in a reasonable time. \\
Although having, today, more qubits than their circuit oriented analogs, quantum annealers are limited to optimisation problems in the form of Quadratic Unconstrained Binary Optimisation (QUBO) problem.
In image segmentation, graph based methods consist in representing the image as a graph, where the nodes represent the pixels and the edges represent the neighbour between pixels. Hence, segmenting the image is equivalent to a graph problem, like coloring problems \cite{graph_coloring} or graph cut \cite{graph_cut}. Among these graph methods, Markov Random Fields (MRF) provide an effective way to model the spatial dependencies in image pixels \cite{mrf_image_kato}. \\
The goal of this paper is to pursue this work by proposing a QUBO formulation for MRF in order to perform unsupervised image segmentation, and compare the performances of our quantum approach with classical results and methods. We also propose an Expectation-Maximisation (EM) \cite{algo_EM} inspired approach to achieve a satisfying unsupervised segmentation of Synthetic Aperture Radar (SAR) images. In the end of the paper, we propose a comparison of our algorithm with non-quantum approaches, and we discuss about its performance and scalability. 

\section{Mathematical Background}

\subsection{Markov Random Fields}

MRF are probabilistic graphical models representing a joint probability distribution over a set of random variables. In an MRF, each node represents a random variable, and the edges between nodes represent statistical conditional dependencies between variables. Let $X$ be a set of observables and $Z$ a set of hidden variables on which one aims to infer. Following the Bayes' inference rule \cite{bayes}, we have :
\begin{equation}
   p(Z = z \mid X = x) \propto p_\theta(X = x \mid Z = z)p(Z = z) \ ,
\end{equation}
with $\theta$ a given set of parameters characterizing the distribution. In the following, we denote by $x = \{x_i\}_{i \in \{1, \dots, n\}}$ the gray scale intensity of a pixel and $z = \{z_i\}_{i \in \{1, \dots, n\}}$ its label. The associated random variables counterparts are denoted by capital letters $X$ and $Z$ respectively. \\
The Maximum A-Posteriori (MAP) $p(Z = z \mid X = x)$ (noted $p(z \mid x)$ in the following for the sake of simplicity) is derived as follows:
\begin{align}
    \ \hat{z} &= \arg \max_{z} p(z \mid x)\ \ , \nonumber \\ 
    &= \arg \max_{z} \prod_{i=1}^n p_\theta(x_i \mid z_i)p(z) \ , \nonumber \\ 
    &= \arg \min_{z} \sum_{i=1}^n - \left(\log p_\theta(x_i \mid z_i)) + \log(p(z))\right) \label{eq:1a} \ .
\end{align}
By identification from \eqref{eq:1a}, we pose $\psi_i(z_i; \theta)$ and $\phi_{i,j}(z_i,z_j)$. For $\phi_{i,j}(z_i,z_j)$, we make the hypothesis that the joint probability distribution $p(z)$ can be decomposed in a product over the pairs of neighbouring nodes \cite{MRF_cliques}, which leads to the following general formulation:
\begin{equation}
    \hat{z} = \arg \min_{z} \sum_{i=1}^n \psi_i(z_i; \theta) +\sum_{\substack{i=1 \\ j \in V_i}}^n \phi_{i,j}(z_i,z_j)
\end{equation}
We note $V_i$ the ensemble of neighbouring nodes of the node $i$. The left term concerns the compatibility of the observable $x$ with the hidden variable $z$. It is the "unary" term of the MRF corresponding to the log-likelihood of node $i$ to be associated with hidden variable $z_i$ knowing its observable $x_i$. The second term is the "pairwise" term of the MRF, which characterizes the compatibility of neighbouring hidden variables. Hence, making the MAP estimation of the MRF is equivalent to find the set of hidden variables $z$ minimizing the sum of both terms, which is NP-Hard \cite{MRF_nphard}. In the literature, classical algorithms as belief propagation \cite{belief_propagation} or variational methods \cite{variational_method} are used to find the optimal sequences of hidden variables. In practice, for a large number of observables, computation time explodes as the number of possible hidden variables sequences exponentially increases \cite{MRF_nphard}.

\subsection{MRF for Image Segmentation}

In the following section, we will consider adapting the MRF model presented above for segmenting a grey-scale image of $N$ pixels in $Q$ distinct classes. The observable $x_i$ corresponds to the intensity of the $i^{th}$ pixel and its label is $z_i$. Pixel intensity takes values from $0$ (black) to $255$ (white). Our input data is the set of pixels $i \in \{1,\dots,N\}$, the set of labels $q \in \{1,\dots,Q\}$ and the set of neighbouring pixels $V_i$ for a pixel $i$. Let's pose the energy function $H(z)$ such as : 
\begin{equation}
    H(z) = \sum_{i=1}^N \psi_i(z_i; \theta) +\sum_{\substack{i=1 \\ j \in V_i}}^N \phi_{i,j}(z_i,z_j) \ . \label{eq:14a}
\end{equation}
The goal is to find a parameterization of the functions $\psi_i(z_i)$ and $\phi_{i,j}(z_i,z_j)$ to make the minimum of $H(z)$ correspond to a satisfying segmentation of the image. Here, $\psi_i(z_i)$ can be interpreted as the function measuring the cost of assigning the label $z_i$ to pixel $i$ knowing its intensity $x_i$. $\phi_{i,j}(z_i,z_j)$ can be interpreted as a function measuring the cost of assigning labels $z_i$ and $z_j$ to the two neighbouring pixels $i$ and $j$. From \eqref{eq:1a} we deduce :
\begin{equation}
    \psi_i(z_i = q; \theta) = -\log(p_\theta(x_i = I_i \mid z_i = q)) \ ,
\end{equation}
with $q \in \{1,\dots,Q\} $ and $I_i \in \{0,\dots,255\}$ the intensity of the pixel $i$. The minimal value of this term is attained for $p(x_i \mid z_i)$ maximal. Hence, this term favors that each pixel is associated to the class he most likely belongs to by only considering its intensity. From \eqref{eq:14a}, we pose the following pairwise term:
\begin{equation}
    \phi_{i,j}(z_i,z_j) = B_{i,j}\delta(z_i,z_j) \label{eq:5a} \ , 
\end{equation}
with $B_{i,j}$ a real positive number and $\delta(z_i,z_j) = 0$ if and only if $z_i = z_j$, $\delta(z_i,z_j) = 1$ else. Literature provides different values for $B_{i,j}$. In the Potts model \cite{potts}, $B_{i,j}$ is defined as a constant value. In other models like Cauchy \cite{cauchy} or Huber \cite{huber} models, $B_{i,j}$ is a function of the intensity of the intensities of $x_i$ and $x_j$. In this paper, for the sake of simplicity, we will consider the Potts model. In the result section, we will discuss the setting of $B_{i,j}$


This intuition behind this formulation comes from the hypothesis that connected pixels are more     likely to have similar labels in a neighbourhood. Hence, the pairwise term favors the fact that all the pixels of the image have the same label, but combined with the unary term, it prevents the labelling of isolated pixels in the wrong class.

\subsection{Expectation Maximisation Algorithm}

The Expectation Maximisation (EM) algorithm is an unsupervised iterative algorithm used to make a Maximum Likelihood Estimation (MLE) \cite{algo_EM} of statistical models parameters with missing data. In our work, we make the hypothesis that, for each region, the distribution of $x_i$ follows some known probability distribution model, and use EM to estimate the probability distributions for the unary term of the MRF. The EM algorithm can be decomposed in two steps : the expectation step (E-step) and the maximisation step (M-step). In the E-step, the algorithm computes the cost function of the problem for a given set of parameters $\theta^t$. This step returns a set a hidden variables and its corresponding cost i.e. energy. The M-step consists in updating the problem parameters, by maximizing the expected log-likelihood computed during the E-step. By alternatively repeating these two steps, the algorithm converges to a local maximum of the likelihood function.\\
Even if demonstrations of the theoretical convergence of the EM algorithm have been achieved \cite{convergence_EM}, EM remains sensitive to the value of $\theta^{0}$ i.e. the value of $\theta^{t}$ at step $0$, which may lead to sub-optimal solutions \cite{init_EM}. However, in our problem, as we do not have the exact ground truth for SAR images, our requirements of quality for the solutions allow us to be satisfied by near-optimal values of the estimated parameters $\hat{\theta}$.

     
\section{Problem Implementation}

As we have seen in section 2.2, we have on one hand MRF which are powerful models for image processing, embedding both unary and pairwise interactions between pixels. On the other hand, EM is a algorithm used for parameter estimation of statistical models, but with a computationally expensive Expectation step (E-step). Previous work shows combination of EM and MRF to obtain satisfying segmentation. \cite{usup_adaptative} proposes an approach where problem parameters are estimated without consideration of their spatial distribution, then executes a MRF to infer the pairwise interaction term. \cite{multilayer_mrf} includes parameters based on the texture of images in addition to the color/intensity observables. In the following sections, we propose two QUBO formulations for the MRF image segmentation. The first one is a two-classes segmentation approach and the second one is a generalized version for $Q$ classes.



\subsection{QUBO formulation for 2-classes segmentation}

Is this section, we propose a QUBO formulation to segment an image in two regions, formulated as a graph cut problem. Graph cut problems are a class of optimization problem aiming to partition a graph into different subsets, such that some cost is minimized \cite{graph_cut}. In the literature, algorithms as Grab Cut \cite{grabcut} or Alpha-Expansion \cite{alpha_expansion} are used to segment images through graph cut methods without any quantum hardware implementation.\\
In this paper, a cut is characterised by two neighbouring pixels $i$ and $j$ having two different labels i.e. with $z_i \neq z_j$. If the cut results in a increase of the cost function value, we speak of penalty. If the cut results in a decrease of the cost function value, we speak of a bonus. \\
Here, we consider an image of $N$ pixels. For a pixel $i$, we note $V_i$ the set of its neighbouring pixels. To each pixel is associated a label $z_i \in \{0,1\}$ and a grey scale intensity $x_i \in \{0,255\}$. Label value $z_i = 1$ implies that the pixel belongs to the "object" class, and $z_i = 0$ implies that it belongs to the "background" class. For each value of $x_i$, we associate the posterior probability $p(x_i \mid z_i = 1)$ and $p(x_i \mid z_i = 0)$ such as $p(x_i \mid z_i = 0) + p(x_i \mid z_i = 1) = 1$. If two pixels are neighbours, there exists an edge between their two respective nodes in the graph.

In order to implement the unary term, we also introduce two ancillary pixels $a$ and $b$ with respective labels $z_a = 1$ and $z_b = 0$. We consider that all pixels of the image are neighbours to these two ancillary pixels. Thereafter, in the graphical expression, we will call the nodes associated to these pixels the "ancillary" nodes. There exists an edge between each ancillary node and image nodes.

Following this formulation, each image node is bound to the nodes of its neighbours in the image plus to the two ancillary nodes. Our goal here is to parameterize the cost of cutting these edges, resulting in the global cost function of the graph cut problem, in the form of QUBOs. In the following sections, we will refer to the different cost function as Hamiltonians, in reference to the quantity defining the energy of a system in quantum mechanics

First, we need to guarantee that $z_a = 1$ and $z_b = 0$. Hence, we pose the linear "constraint" Hamiltonian:
\begin{equation}
    h_A(z_a, z_b) = (z_b - z_a) \label{eq:2a}
\end{equation}
$h_A(z_a, z_b)$ is minimal if $z_a = 1$ and $z_b = 0$.  
We also define the "cut" Hamiltonian as follows :
\begin{equation}
    \delta(z_i, z_j) = z_i + z_j - 2z_iz_j
\end{equation}
This Hamiltonian is equal to $0$ if $z_i = z_j$, and equal to $1$ if $z_i \neq z_j$. Then, we define the "unary" Hamiltonian as follows :
\begin{equation}
    H_U(z, z_a, z_b; \theta) = \sum_{i=1}^N \delta(z_i, z_a)\log(p_\theta(x_i \mid z_a)) + \sum_{i=1}^N \delta(z_i, z_b)\log(p_\theta(x_i \mid z_b)) \label{eq:3a}
\end{equation}
If $z_i \neq z_a$, then $\delta(z_i, z_a) = 1$ and $\delta(z_i, z_a)\log(p(z_a \mid x_i)) < 0$ (same for $z_b$). Hence, the pixel is labelled as the class it has the highest probability to belong. The global minimum of this Hamiltonian corresponds to each pixel being labelled to the class it has the highest probability to belong, which satisfies the formulation of 2.2.
For the pairwise term, we pose the following "pairwise" Hamiltonian :
\begin{equation}
    H_P(z) = \sum_{i=1}^N \sum_{j \in V_i} \delta(z_i, z_j) \label{eq:4a}
\end{equation}
In this expression, a penalty of value $1$ is applied if $z_i \neq z_j$. From \eqref{eq:2a}, \eqref{eq:3a} and \eqref{eq:4a}, we pose the "problem" Hamiltonian as follows :
\begin{equation}
    H(z, z_a, z_b; \theta) = H_U(z, z_a, z_b; \theta) + \lambda_PH_P(z) + \lambda_Ah_A(z_a, z_b) \label{eq:6a}
\end{equation}
With $\lambda_P$ and $\lambda_A$ two positive real multipliers. From \eqref{eq:5a}, we pose $\lambda_P = B$ a positive real constant.


For $\lambda_A$, we have to ensure that it is always favorable, in term of value of the cost function, to respect that $z_a = 1$ and $z_b = 0$ for all labelling $z$. To do so, let's consider the extreme scenario where all the pixels of the image are labelled in the "object" (resp. "background") class, and all pixel have an equiprobability of belonging to class "object" or "background". Let's suppose $z_a = 1$ and $z_b =0$. Following the above conditions, $\forall i, z_i = z_a$ and $p_\theta(x_i \mid z_a) = p_\theta(x_i \mid z_b) = 0.5$, Hence, we have here the maximum possible minimal value of \eqref{eq:3a}. In that case, setting $z_b = 1$ would apply a bonus of $Nlog(0.5)$ to the cost function. In order to guarantee that this configuration will never be favored, we have to set $\lambda_A > -Nlog(0.5)$. Hence, $\forall z$, with $z_a = z_b$, $H_U(z, z_a, z_b; \theta) + \lambda_Ah_A(z_a, z_b) > 0$, which ensures that any sequence minimizing $H(z, z_a, z_b; \theta)$ also minimizes $h_A(z_a, z_b)$.

\subsection{QUBO formulation for Q-classes segmentation}

In this part, we extend the formulation presented in 3.2 for $Q$ classes. Lets consider an image of $N$ pixels that we want to divide into $Q$ disjoint subsets. In order to encode the $Q$ possible labellings for each pixel, we pose the vector $z' \in \{0,1\}^{NQ}$, and we note $z'_i$ each bloc of length $Q$ that is assumed to have only one non-null value. In the following, we define the function $\psi(i,q) = (i-1)Q+(q-1)$ such as $z_{\psi(i,q)}$ is the $i^{th}$ bloc of $z$ having its non-null value at the $q^{th}$ index. This method is called one-hot encoding, and we will consider that $z_{\psi(i,q)} = 1$ implies that pixel $i$ is labelled $q$. The rest of notations remains unchanged.

In order to ensure one-hot encoding for all, we define the "one-hot" Hamiltonian as follows :
\begin{equation}
    H_{OH}(z) = \sum_{i=1}^N \sum_{q=1}^Q \left(-z_{\psi(i,q)} + 2 \sum_{r<q}^Q z_{\psi(i,q)}z_{\psi(i,r)}\right) \label{eq:7a}
\end{equation}
As in the binary case, we have to define a single ancillary pixel $\alpha$ such as $z_\alpha = 1$. To ensure its value, we pose the following "constraint" Hamiltonian :
\begin{equation}
    H_{A'}(z_\alpha) = -z_\alpha \label{eq:8a}
\end{equation}

The "unary" Hamiltonian for the Q-classes is defined as follows :
\begin{equation}
    H_{U'}(z, z_\alpha; \theta) = \sum_{i=1}^N \sum_{q=1}^Q \delta(z_{\psi(i,q)}, z_\alpha)\log(p_\theta(x_i \mid z_{\psi(i,q)}))
    \label{eq:9a}
\end{equation}
Following a similar principle as the unary Hamiltonian of 3.2, the goal is to maximize the sum of bonuses, by cutting the most negative edges. Here, one-hot encoding imposes that there are $Q-1$ cuts between nodes corresponding $z_{\psi(i,q)} \ \forall q$ and $z_\alpha$ for a given $i$. The optimal cut is achieved when the $Q-1$ most negative edges are cut, which corresponds to the $Q-1$ lowest $p_\theta(x_i \mid z_{\psi(i,q)})$. Hence, the only non-cut occur for the maximal value of $p_\theta(x_i \mid z_{\psi(i,q)})$ which corresponds to maximum likelihood term in 2.2. \\
The "pairwise" Hamiltonian for the Q-classes is defined as follows :
\begin{equation}
    H_{P'}(z) = \sum_{i=1}^N \sum_{j \in V_i} \sum_{q=1}^Q \frac{\delta(z_{\psi(i,q)}, z_{\psi(j,q)})}{2} \label{eq:10a}
\end{equation}
Considering that one-hot encoding is respected, if pixels $i$ and $j$ have the same class, then $z_{\psi(i,q)}= z_{\psi(j,q)} \ \forall q$ and no penalty is applied. \\
From \eqref{eq:7a}, \eqref{eq:8a}, \eqref{eq:9a} and \eqref{eq:10a}, we deduce the global QUBO formulation for the Q-class segmentation:
\begin{equation}
    H'(z, z_\alpha; \theta) = H_{U'}(z, z_\alpha; \theta) + \lambda_{P'}H_{P'}(z) + \lambda_{OH}H_{OH}(z) + \lambda_{A'}h_{A'}(z_\alpha) \label{eq:13a}
\end{equation}
With $\lambda_{P'}$, $\lambda_{OH}$ and $\lambda_{A'}$ three positive real Lagrange multipliers. For $\lambda_{P'}$, as the pairwise interactions follow the same hypothesis, we set $\lambda_{P'} = \lambda_{P} = B$. For $\lambda_{A'}$, by following the same reasoning as in 3.1, we can again consider an extreme scenario. In the Q-class state, an equiprobability of belonging to each class implies that $p(z_{\psi(i,q)} \mid x_i) = 1/Q \ \forall q$. Because there are $Q-1$ cuts per pixel, the total number of cuts between the image nodes and the ancillary node is $N(Q-1)$. Note that this number of cuts is independent from the labelling. From the results of 3.2, we deduce that we have to set $\lambda_{A'} > -N(Q-1)log(1/Q)$ in order to guarantee that the constraint is always respected for an optimal value of $H'$.

For the minimal value of $\lambda_{OH}$, we have to consider the situation where neighbouring pixels $i$ and $j$ have different labels $p$ and $q$. In this situation, a penalty of $B$ is applied, which corresponds to the sum of two cuts. The first cut is between $z_{\psi(i,p)} = 1$ and $z_{\psi(j,p)} = 0$ and the second between $z_{\psi(i,q)} = 0$ and $z_{\psi(j,q)} = 1$. In this configuration, in order to guarantee that it is never favorable to set $z_{\psi(i,p)} = z_{\psi(j,p)}$ and/or $z_{\psi(i,q)} = z_{\psi(j,q)}$, we have to set $\lambda_{OH} > \frac{\lambda_{P'}}{2}$.


\section{EM inspired approach for parameter estimation}

In this part, we address the parameter estimation problem of the unary term of the MRF. For the sake of generality, the formulation we propose is based on the Q-classes formulation of 3.2, as setting $Q=2$ is equivalent to a binary class formulation of 3.1. In this section, we propose an Expectation Maximization algorithm \cite{algo_EM} to estimate the parameters of each distribution, then generalize it to any distribution and infer a general formulation for \eqref{eq:10a}. \\
In this section, we describe the algorithm used to obtain an unsupervised segmentation of an image of $N$ pixel in $Q$ segment. Pixel intensities with one segment are assumed to be a sample of a normal distribution with parameters  $\theta_q = (\sigma_q, \mu_q)$. For each step $t$ of the EM algorithm, we note $\theta_q^t$ the distribution parameters of segment $q$ at step $t$ and we pose $\theta^t = \{\theta_1^t, \dots, \theta_Q^t\}$. The initial distribution parameter of the algorithm is noted $\theta^0$. \\
The literature underlines the importance of well initializing $\theta_q^0$ in order to efficiently converge to a global minimum \cite{init_EM}. In our approach, we used a $k$-means algorithm \cite{k_means} in order to initialize the values $\theta_q^0$ for $Q$ clusters. This algorithm provides a value of $\mu_q^0$ for each class, and a single value of variance $\sigma$ for all clusters. For the sake of simplicity and computation cost, we consider that setting $\sigma_q^0 = \sigma \ \forall q$ is a satisfying initial variance value for all segments. \\
For the E-step, we compute an expectation of the MRF for the set of parameters $\theta^t$. As a result, we obtain $Q$ subsets of pixels intensity values $x_q^t$, one for each segment. For the M-step, the goal is to find the set of parameters $\theta^{t+1}$ such as \eqref{eq:13a} attains new minimum. If $p_\theta(x_i \mid z_i)$ follows a Gaussian model, this minimization is explicit \cite{gauss_explicit}, otherwise, minimization has to be performed numerically \cite{weib_implicit}. In order to evaluate if convergence is reached, we compute 
\begin{equation}
    \Delta^{t} = \lVert L(\theta^{t+1}) - L(\theta^t) \rVert \ ,
\end{equation}
with $L(\theta^t) = H'(z,z_a ; \theta^t)$ i.e. the equation \eqref{eq:13a} parameterized by $\theta$. If $\Delta^{t}$ is least than some predefined threshold $\delta > 0$, we consider that we have reached convergence, an retains the set of labels $z^t$ obtained during the E-step. Else, $t \leftarrow t+1$ and the algorithm loops back to a new E-step.


\section{Adaptation for SAR image processing}

In this part, we consider adapting the above formulations to SAR images. In the literature, previous work assimilates the distribution of values of pixels in SAR images segments to a Weibull distribution \cite{SAR_weibull, SAR_weibull_2}, which is a special case of the general gamma distribution (GGD). The GGD has the following probability density function :
\begin{equation}
    f(x;\lambda,k,p,t) =
    \begin{cases}
     \ \frac{(p/\lambda^k)(x-t)^{k-1}}{\Gamma(k/p)}e^{-((x-t)/\lambda)^p} \text{ if }  x > t \\
     \ 0 \text{ else}
    \end{cases}
    \nonumber
\end{equation}
With $k,p,\lambda$ and $t$ positive real numbers. In the above formulation, $k$ and $p$ are the shape parameters, $\lambda$ is the scale parameter and $t$ is the location parameter. In the case of a 2 parameters Weibull distribution, $k = p$ and $t=0$. Straightforward calculations lead to the following distribution:
\begin{equation}
    g(x;\lambda,k) =
    \begin{cases}
     \ \frac{k}{\lambda}\left(\frac{x}{\lambda}\right)^{k-1} e^{-(x/\lambda)^k} \text{ if } x > 0\\
     \ 0 \text{ else} 
    \end{cases}
    \label{eq:speckle_modif}
\end{equation}
Unlike the Gaussian distribution, there is no explicit method to estimate the parameters of a Weibull distribution. Hence, for the M-step of the MRF algorithm described in section 4.1, we will rely on basic numerical optimization to update our problems parameters for each class. 


\section{Results}

\subsection{Experimental context}

In this section, we propose and discuss the results obtained on the D-Wave adiabatic quantum computer \cite{D-Wave}. D-Wave is a Canadian company which proposes access to their quantum devices through a cloud based service called D-Wave Leap. The device we use in the following is the D-Wave Advantage quantum computer, having 5760 available qubits and 15 couplers per qubits. The number of couplers quantifies the number of interconnections available between physical qubits. In the QUBO formulations presented above, a product between two variables implies there is a coupler between their two corresponding qubits. As this number of products can exceed 15 in our formulations, the embedding function provided by D-Wave libraries binds qubits with each other, creating logical qubits from multiple physical qubits. In the computation, logical qubits act as a single variable. Consequently, due to this embedding constraint, the actual limitation of problem size that can be implemented on the quantum machine is below the theoretical limitation ($N^2 + 2 < 5760$ for example for the bi-classes formulation). Recent publications from D-Wave present hardware having 7800+ qubits available and more interconnections. However, in the following, we will consider current limitations of D-Wave hardware at 5760 qubits and 15 interconnections.

In order to compare our results, we also used a simulated annealing library provided by D-Wave called d-wave neal. We use this tool in order to implement larger instances of the problem which cannot be implemented in the quantum hardware. More precisely, we chose to simulate our results through the d-wave neal routines in order to segment images of a sufficient size to highlight the performances of our approach. For 2 classes segmentation, the maximum size of instances we can implement on the quantum computer is $(22,22)$ with the binary formulation and $(15,15)$ for the multi class formulation (with $Q=2$). For  $Q = 3$, the maximum image size is $(9,9)$. Through our experiments, it appeared that the major limitation in term of embedding was the maximal number of connections between variables. As we had no access to D-Wave's embedding routines in order to find better formulations to our problem regarding the hardware constraints, we did not focus on finding an optimal embedding in our work. In comparison, d-wave neal enabled us to obtain results for larger images, up to $(100,100)$ for 2 classes segmentation and $(80,80)$ for 3 classes segmentation, with a limitation in term of memory on our classical devices.

Unfortunately, as we write these lines, d-wave neal does not provide a quantum noise simulator, resulting in over-optimistic results. Consequently, our estimations of algorithm performances will be based on an extrapolation of performances obtained for small images. In order to put the results in perspective, the following sections propose a quantitative and qualitative comparison with a region smoothing algorithm presented in \cite{region_smoothing}. In this paper, the authors compare their approach with other unsupervised segmentation algorithms for both gaussian noisy and SAR images. Their results appear to be significantly better in computation time and accuracy on toy examples compared to fuzzy approaches. We also tested the fuzzy clustering algorithm proposed in \cite{fuzzy_clustering}, which gave satisfying results for gaussian noisy images but did not show good results on SAR images. 

\begin{figure}[!h]
    \centering
    \captionsetup{justification=centering,margin=1cm}
    \includegraphics[scale=0.45]{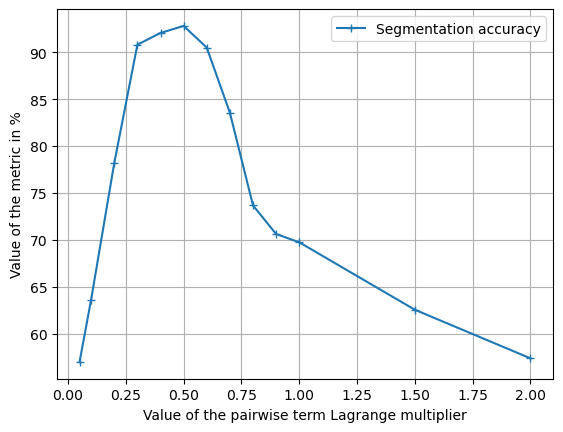}
    \caption{Variation of the accuracy depending on the value of $\lambda_{P}$. These results are obtained by simulating the multi-class segmentation approach on a (40, 40) image with $Q = 4$.}
    \label{fig:acc_pairwise}
\end{figure}

Concerning the weight $B$ of the pairwise term, search heuristics presented in \textbf{Fig. \ref{fig:acc_pairwise}} have shown that best values were obtained for $\lambda_P$ around $0.5$, which corroborates the results presented in \cite{potts, pairwise_ponderation}. Moreover, it appeared in our tests that fine tuning the value of $\lambda_P$ as a function of $Q$ would improve the quality of segmentation. Hence, in all the following test, we set $\lambda_p = 0.6$, $\lambda_p = 0.5$ and $\lambda_p = 0.35$ for resp. $Q=3$, $Q=4$ and $Q=5$.

In the following, segmentation accuracy is defined as the percentage of well-labelled pixels. The computation time for a single sample (the execution time on the quantum annealer) is the sum of the delay time, the annealing time and the readout time. Delay time is fixed at $21 \mu s$, and consist in initializing the system in an equiprobability of obtention of any sequence. Then, the annealing time represents the convergence time from the initial state to the final state. The latter is defined by the QUBO, and its minimal energy configurations correspond to solutions minimizing the criterion. Previous work shown that the annealing time logarithmically scales with the number of variables \cite{temps_annealing}. In our work, for the sake of simplicity, we will consider an annealing time of $200 \mu s$ regardless the number of variables. Concerning the readout time, it logarithmicaly scales with the number of variables, for a value in the order of $100 \mu s$.\\
Due to the inherent error of quantum computation, we consider repeating the annealing, and then keep the best result i.e. the set of labels with the minimal energy among all trials. Hence, repeating the number of trials increases the chances to obtaining an optimal solution. In section 6.4, we will discuss the settings to increase the probability of obtention of optimal solutions whilst minimizing the hardware related error. \\

\subsection{Results for Gaussian distributions}

In this section, we consider segmenting images with an additional Gaussian noise. For binary and multi-classes segmentation, we will respectively use formulations \eqref{eq:6a} and \eqref{eq:13a} as the E-step of the EM algorithm. Due to the variable queue time to access the D-Wave hardware, we consider that the computing time of the E-step is equivalent to the computation time for a single sample times the number of samples. Hence, the total computation time is equivalent to the initialization time (problem embedding + k-means algorithm), plus the number of steps of the EM-algorithm times the computation time of the E-step and the M-step. In order to get a better result and limit the impact of error at the end of the algorithm, once convergence is reached, we re-compute the E-step for a greater number of samples, and keep the labelling corresponding to the minimum energy as our result.

\begin{figure}[!ht]
    \centering
    \captionsetup{justification=centering,margin=1cm}
    \includegraphics[scale=0.4]{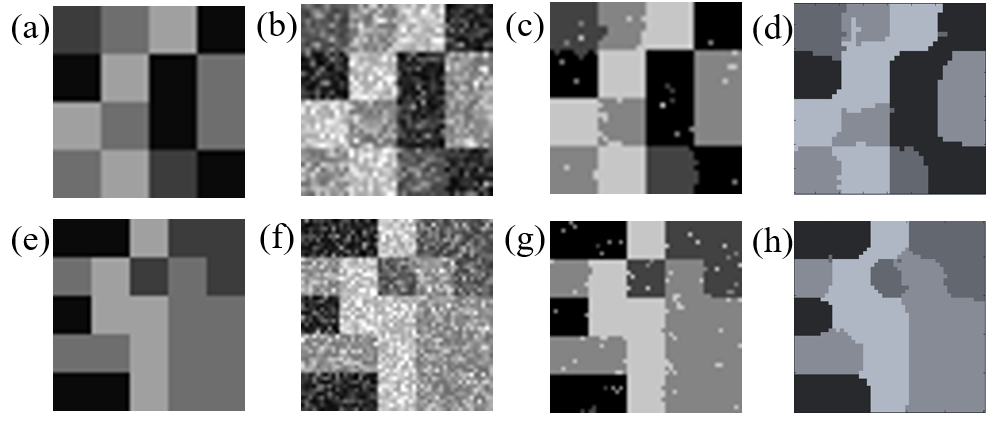}
    \caption{Qualitative results of the EM algorithm for images with additional Gaussian noise: (a) and (e) corresponds the original image, (b) and (f) corresponds to the original image with the addition of a gaussian noise, (c) and (g) correspond to the segmentation obtained with the simulated annealing approach. (d) and (h) corresponds to the segmentation obtained with the region smoothing algorithm \cite{region_smoothing}}
    \label{fig:qualitive_results}
\end{figure}

\textbf{Fig. 2} presents qualitative results for generated images with Gaussian noise using our approach and the region smoothing algorithm proposed in \cite{region_smoothing}. In \textbf{Table 1}, quantitative results  are shown for different generated checkerboards with $Q = 3$ to $Q = 5$ classes. It presents the improvement of accuracy provided by the EM algorithm, as the parameters of each Gaussian corresponds to the ones of the generated image. For \textbf{Table 1}, \textbf{Table 2} and \textbf{Fig. 2}, we present results obtained with the d-wave neal sampler, as hardware embedding constraints (especially one-hot encoding) are not matched for a high number of pixels and sequences respecting the one-hot encoding constraints for all pixels are rarely obtained for large instances (nb. of logical variables $> 250$). Nevertheless, for the binary formulation, as no one-hot encoding is required, embedding errors results in a loss of accuracy, without violating the constraints. For all the test on d-wave neal, the number of samples to $100$ samples, the convergence criteria to $\delta = 5$, the annealing time to $250 \mu s$ and the maximal number of epochs to $T = 30$.

\begin{table}[!ht]
    \centering
    \begin{tabular}{ |p{1.5cm}|p{0.8cm}|p{0.8cm}|p{1.2cm}||p{2.3cm}|p{2.3cm}|p{2.3cm}| }
         \hline
         \multicolumn{7}{|c|}{Quantitative results for Gaussian noise image segmentation} \\
         \multicolumn{4}{|c||}{Problem parameters} & \multicolumn{3}{|c|}{Accuracy at a given epoch} \\
         \hline
         Img. size & Q &$\sigma$&nb. var. &T=1 &T = 10 &T = 30 \\
         \hline
            (40,40)   & 3   & 50  & 6401 &92.17\%&  96.08\% & 97.56\% \\
            (40,40)   & 4   & 25  & 6401 &96.66\%&  97.69\% & 99.60\%\\
            (40,40)   & 4  & 50  & 6401 &78.28\%&   85.38\% & 91.38\% \\
            (40,40)   & 5  & 50  & 8001 &82.47\%&   89.73\% & 94.75\% \\
            (50,50)   & 3  & 25  & 7501 & 93.30\%&  97.21\% & 98.50\% \\
            (50,50)   & 3  & 50  & 7501 & 84.89\%&  92.92\% & 96.32\% \\
            (50,50)   & 4  & 50  & 10001 & 80.48\%&  91.80\% & 95.76\% \\
            (50,50)   & 5  & 50  & 12501 & 77.57\%& 86.41\% &91.61\%\\
         \hline
    \end{tabular}
    \\
    \caption{Average accuracy for multi-class segmentation for images with added Gaussian noise. In the above, $Q$ corresponds to the number of class and $\sigma$ corresponds to the variance of normal distribution of the Gaussian noise. Nb. var. corresponds to the number of (logical) qubits required to implement the problem. $T$ represents the number of iterations.}
    \label{tab:my_label}
\end{table}

Quantitative results presented in \textbf{Table 1} underline the advantage of using the EM Algorithm to improve the quality for the segmentation. It appears that this improvement is proportional to the number of classes, the size of the image and the value of $\sigma$. 
Regardless the embedding constraints, computing the E-step for 100 samples on the quantum annealer would take at most $ \approx 100 \times 200 \mu s = 0.02s$ plus $\approx 0.2s$ to update the cost function and re-implement it on the quantum hardware. For the M-step, the approximate computation time is $0.05s$. These results leads to an approximate computation time of $ \approx 0.3s$ per epoch, which can be even more reduced by a faster classical computation. This duration logarithmicaly scales with the number of variables, as the optimal annealing time and readout time logarithmically scales with the number of qubits. As a comparison, d-wave neal sampler has a computation time of 5.3 seconds per epoch (image of size (40,40) and $Q = 4$) for the E-step (cost function update + computation) for $100$ samples, which is consistent with the results of \cite{em_hard_1, em_hard_2} obtained with non-quantum approaches.

\begin{table}[!ht]
    \centering
    \begin{tabular}{ |p{1.5cm}|p{0.5cm}||p{1.5cm}|p{1.4cm}||p{1.5cm}|p{1.4cm}||p{1.5cm}|p{1.4cm}| }
         \hline
         \multicolumn{8}{|c|}{Comparison of performances with classical approaches} \\
         \multicolumn{2}{|c||}{Pb. parameters} & \multicolumn{2}{|c||}{Simulated annealing} & \multicolumn{2}{|c||}{Region smoothing} &\multicolumn{2}{|c|}{Fuzzy clustering}\\
         \hline
         Img. size & Q &Acc. &t &Acc. & t &Acc. &t   \\
         \hline
            (40,40)   & 3    &97.56\%&  152.80 & 98.94\% &0.14& 95.19\% &0.66\\
            (40,40)   & 4      &96.69\%  &167.34 & 96.28\% &0.17& 90.56\% &0.82\\
            (40,40)   & 5     &94.75\%  &183.93 & 91.06\% &0.20& 85.06\% &1.01\\
            (50,50)   & 3     &96.32\% &178.18 & 97.16\% &0.18& 94.02\% &0.91\\
            (50,50)   & 4    &95.76\%&  199.78 & 94.72\% &0.22& 88.68\% &1.18\\
            (50,50)   & 5   &91.61\%&   224.71 & 86.96\% &0.27& 81.19\% &1.39\\
         \hline
    \end{tabular}
    \\
    \caption{Comparison of average computation time and accuracy between simulated annealing and other non-quantum approaches proposed in \cite{region_smoothing} and \cite{fuzzy_clustering}. Simulated annealing results are obtained by running our algorithm on 30 epochs using d-wave neal for the E-step. Notation $t$ corresponds to the computation time in seconds.}
    \label{tab:table_time}
\end{table}

In \textbf{Table 2}, we compare our results with region smoothing and fuzzy clustering algorithms resp. proposed in \cite{region_smoothing} and \cite{fuzzy_clustering}. Simulated annealing results obtained with d-wave neal underline the computation cost of sampling a MRF with classical approach. The corresponding computation time for each instance would be $\approx 9$ seconds ($30 \times 0.3$ seconds, average computation time mentioned above), time that could be consequently reduced by a faster computation of non-quantum calculus. For each epoch, sampling the MRF only takes $0.02s$ of quantum computation for $100$ samples, instead of more than $100s$ in simulated annealing. This implies that processing larger images would not significantly increase the computation time. Moreover, assuming that quantum annealing would provide the same results as simulated annealing, for the largest instances, the algorithm provides better results in terms of accuracy than the two other approaches. These latter assessments lead us to think that our algorithm would be relevant for segmenting large SAR images, as described in the introduction.

\subsection{Results for SAR images}

\begin{figure}[!h]
    \centering
    \captionsetup{justification=centering,margin=1cm}
    \includegraphics[scale=0.6]{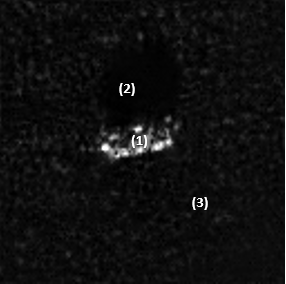}
    \caption{Example of MSTAR SAR image: (1) corresponds to the vehicle, (2) corresponds to the shadow of the vehicle, (3) corresponds to the (noisy) background of the image.}
    \label{fig:example_SAR}
\end{figure}

In this section, we present results on SAR images with the adapted parameterization \eqref{eq:speckle_modif} described in section 5. For these tests, we will only provide qualitative results, as ground truth is not available. First results are proposed on cropped images of the MSTAR dataset \cite{MSTAR}, composed of various military vehicles SAR images similar to \textbf{Fig. \ref{fig:example_SAR}}. As evoked in the introduction, datasets composed of MSTAR-like segmented images can be used to train ATRt classifiers \cite{ATR_SAR}. Hence, our goal is to subdivide these images in 3 segments, corresponding to the vehicle, its shadow and the rest of the image. The object corresponds to a connected area composed of high intensity pixels and the shadow corresponds to a connected low intensity pixels area next to the object. For the chosen MSTAR images, we consider that a satisfying segmentation is obtained if the object and shadow segments are both connected areas next to the center of the image, without including background labelled pixels.

As we aim to segment the image into 3 classes, the E-step of the EM algorithm will be executed with the \eqref{eq:13a} formulation, adapted to SAR images. Concerning the initialization of the EM algorithm, k-means and Weibull mixture models did not provide an accurate set of parameters, which led to non-satisfying segmentation. Hence, we chose to initialize our model with a thresholding method. Thus, we consider that pixels with intensity of $x_i \leq 7$ most likely belong to the "shadow" class and pixels with an intensity $x_i \geq 20$ most likely belong to the "target" class. For other intensity values, we consider that they most likely belong to the "background" class.

\begin{figure}[!h]
    \centering
    \captionsetup{justification=centering,margin=1cm}
    \includegraphics[scale=0.70]{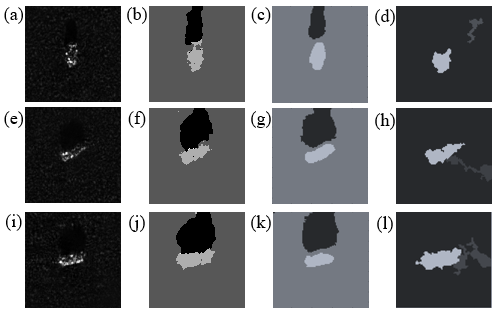}
    \caption{Qualitative results of the EM algorithm on M-STAR SAR images. (a/e/i) correspond to the original image. (b/f/j) correspond to the results obtained with our approach. (c/g/k) corresponds to the results obtained with the region smoothing algorithm. (d/h/l) corresponds to the results obtained with the fuzzy clustering algorithm.}
    \label{fig:qualitative_SAR_image}
\end{figure}

In \textbf{Fig. \ref{fig:qualitative_SAR_image}}, we present results obtained with the d-wave neal sampler on some MSTAR dataset images, and compare its performances with the region smoothing \cite{region_smoothing} and fuzzy clustering \cite{fuzzy_clustering}. In this figure, rightmost images, obtained with the fuzzy clustering algorithm, do not present a satisfying segmentation of SAR images, as the shadow region is associated to the background. On images obtained with the region smoothing algorithm, target and shadow regions are well located but are not connected and roundish, which may complicate the recognition of the target.

\begin{figure}[!h]
    \centering
    \captionsetup{justification=centering,margin=1cm}
    \includegraphics[scale=0.55]{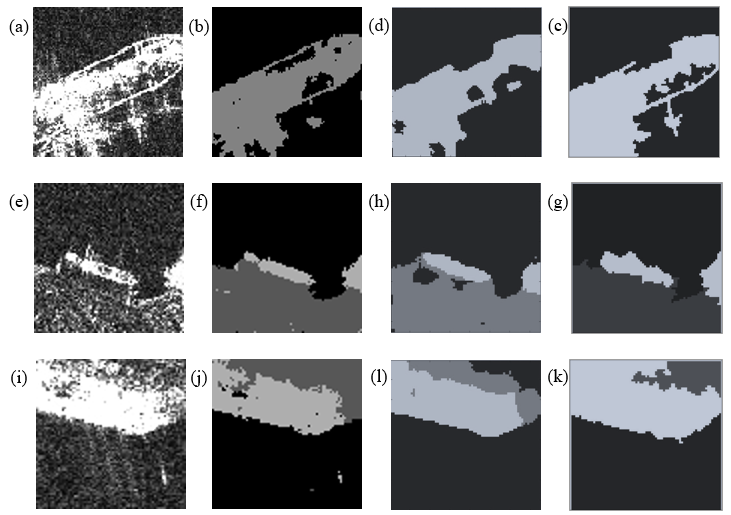}
    \caption{Qualitative results of the EM algorithm on SSDD dataset images. (a/e/i) correspond to the original image. (b/f/j) correspond to the results obtained with our approach. (c/g/k) corresponds to the results obtained with the region smoothing algorithm. (d/h/l) corresponds to the results obtained with the fuzzy clustering algorithm.}
    \label{fig:qualitative_SAR_image_non_MSTAR}
\end{figure}

In \textbf{Fig. \ref{fig:qualitative_SAR_image_non_MSTAR}}, we present the results obtained with the d-wave neal sampler on the SAR Ship Detection Dataset (SSDD) \cite{SSDD_dataset}. As in the previous figure, we compare the performances of our approach with the region smoothing \cite{region_smoothing} and fuzzy clustering \cite{fuzzy_clustering} algorithms. Segmentations b/c/d are obtained on a $(100,100)$ image (a) with two classes (corresponding to the ship and the sea). For the two other segmentations, we consider $(75,75)$ images (e and i). Those images are segmented in 3 classes ranked from the brightest to the darkest: the ship, the land and the sea. On these figures, we highlight the efficiency of our algorithm for removing noise and artifacts on lower-quality SAR images (in comparison to MSTAR dataset images) whilst distinguishing each class with a satisfying accuracy. As in \textbf{Fig. \ref{fig:qualitative_SAR_image}}, evaluated classical approach present less satisfying results, by being less accurate in the preservation of the shape of the ship and the delimitation between the land and the two other classes.

For the same reason as in section 6.2, embedding constraints and hardware error greatly limits the size of the instances we can implement on the quantum annealer. Due to the non relevance of segmenting small images (of maximum size $(15,15)$), we chose to only present our results obtained on d-wave neal as a realistic expectation of  future quantum hardware performances, especially in terms of embedding capability and error tolerance.

\subsection{Discussion on optimal parameters and limits of the model}

In the previous sections, our results underlined the current limitation of quantum technologies regarding the error rate and the embedding constraints. Nevertheless, we can already discuss on optimal parameters for the model as the value of the weighting of the constraints, and also optimal parameter for the computation, as the annealing time.

\begin{figure}[!h]
    \centering
    \captionsetup{justification=centering,margin=1cm}
    \includegraphics[scale=0.45]{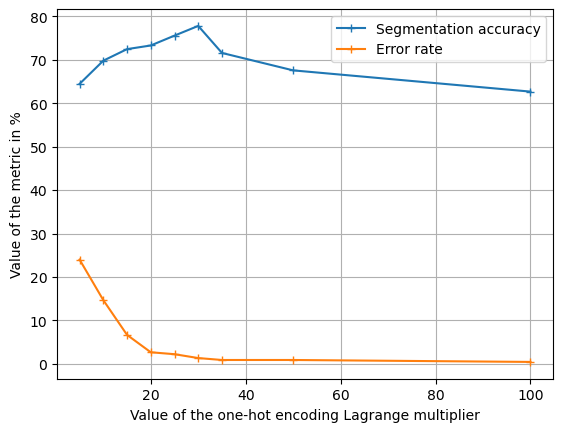}
    \caption{Variation of the segmentation accuracy and the percentage of pixels not respecting the one-hot encoding constraint depending on the value of $\lambda_{OH}$. These results are obtained by executing the multi-class segmentation approach on a $(15,15)$ image with 2 classes, for $100$ samples and an annealing time of $50 \mu s$.}
    \label{fig:err_acc_graph}
\end{figure}

In \textbf{Fig. \ref{fig:err_acc_graph}}, we can notice that increasing the value of $\lambda_{OH}$ exponentially reduces the number of solutions not-respecting the constraint. Nevertheless, for high values of $\lambda_{OH}$, we can notice a reduction of the segmentation accuracy. This reduction is due to the increased energy of the potential barriers in the quantum annealer between two solutions respecting the one-hot encoding constraint. As quantum annealers use tunneling effect to converge to the global optimum, increasing the energy of these barriers reduces the probability of the system to converge to a lower minimum \cite{D-Wave, reflexion_annealing}. Consequently, even if all tested values of $\lambda_{OH}$ respect the requirements presented in section 3.1, its value has to be judiciously chosen to limit the error rate whilst maximizing the segmentation accuracy. For error labelled pixels, completions methods based on the labels of nearest neighbours can be applied to achieve a satisfying segmentation of the image, without loosing the quantum speedup advantage.

\begin{figure}[!h]
    \centering
    \captionsetup{justification=centering,margin=1cm}
    \includegraphics[scale=0.45]{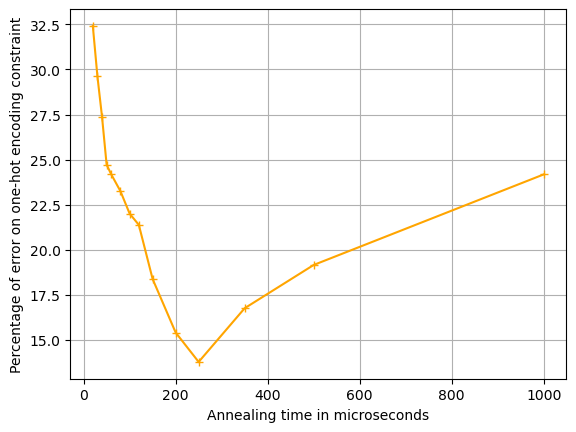}
    \caption{Variation of the percentage of pixels not respecting the one-hot encoding constraint depending on the annealing time. These results are obtained by executing the multi-class segmentation approach on a $(15,15)$ image with 2 classes, for $100$ samples and $\lambda_{OH} = 5$.}
    \label{fig:err_time_graph}
\end{figure}

Concerning the annealing time, \textbf{Fig. \ref{fig:err_time_graph}} also shows a correlation between the error rate and the annealing time, which corroborates the results of \cite{reflexion_annealing}. On this figure, we have chosen a value of $\lambda_{OH}$ such that the impact of the annealing time on the number of errors is clearly visible (same for \textbf{Fig. \ref{fig:err_acc_graph}}). Optimal annealing time is around $250 \mu s$ which is usually a good value \cite{reflexion_annealing} confirmed by our experiments. \\
Disregarding the error of the quantum hardware, larger instances may be implemented by duplicating some variables with a lot of connections. Even if D-Wave embedding functions already binds logical variables to multiple physical qubits, these generic functions may not provide an optimal embedding. Hence, including the optimal embedding in the QUBO formulation could increase the number of implementable instances, at the cost of additional ancillary variables. \\
Moreover, one major hypothesis of the model is the number of classes in the image. In practice, for a SAR image, there is no guarantee that a vehicle is present on the image. Then, the image should have to be segmented in a single "background" class to maximize the accuracy, whilst the EM algorithm will try to find 3 distinct classes. Hence, it could be accurate to use the EM algorithm to find optimal parameters on reference images (with 3 classes) in order to have an initial guess of the classes parameters, then execute the MRF with fixed parameters and $Q = 3$ on all the images of the dataset, including images without vehicles. For similar noise parameters, this approach would achieve a satisfying segmentation of all images, including images without vehicle. 
Another method could be to reduce the number of classes during the EM algorithm process. At a given epoch, if the number of pixels labelled in a certain class is below a predefined limit, we could remove this class and proceed the algorithm for a reduced number of classes. However, this method appears to be highly error sensitive, as a single irrelevant epoch could lead to a definitive removal of a class.

\section{Conclusion}

In conclusion, our work proposes a new hybrid quantum-classical method for image segmentation, taking profit of the quantum annealing speedup to perform fast segmentation of SAR images. We also present our first qualitative and quantitative results on the D-Wave quantum annealer and evaluate the scalability of our method regarding the current hardware constraints. Moreover, we propose a discussion on the optimal parameters for the quantum computation, especially on the weighting of constraint cost functions and the annealing time. We conclude that quantum approaches for image segmentation are promising, but improvements of current hardware limitations are required, especially concerning the error rate and the embedding capability.\\
To complete this paper, further studies could be done on other hardware related parameters, such as optimal embedding. Concerning the hardware, as our formulation is compatible with any quantum annealers, our method could be tested on other quantum computers, as the Pasqal machine \cite{pasqal}, which uses spatially arranged cold atoms to perform quantum annealing. 
Our work can also be extended to other image processing problems useful in the radar domain, as edge detection or denoising.

\bibliographystyle{IEEEtran}
\bibliography{sample}

\begin{thebibliography}{10}
\providecommand{\url}[1]{#1}
\csname url@samestyle\endcsname
\providecommand{\newblock}{\relax}
\providecommand{\bibinfo}[2]{#2}
\providecommand{\BIBentrySTDinterwordspacing}{\spaceskip=0pt\relax}
\providecommand{\BIBentryALTinterwordstretchfactor}{4}
\providecommand{\BIBentryALTinterwordspacing}{\spaceskip=\fontdimen2\font plus
\BIBentryALTinterwordstretchfactor\fontdimen3\font minus \fontdimen4\font\relax}
\providecommand{\BIBforeignlanguage}[2]{{%
\expandafter\ifx\csname l@#1\endcsname\relax
\typeout{** WARNING: IEEEtran.bst: No hyphenation pattern has been}%
\typeout{** loaded for the language `#1'. Using the pattern for}%
\typeout{** the default language instead.}%
\else
\language=\csname l@#1\endcsname
\fi
#2}}
\providecommand{\BIBdecl}{\relax}
\BIBdecl

\bibitem{Intro_SAR}
\BIBentryALTinterwordspacing
T.~P. AGER, ``An introduction to synthetic aperture radar imaging,'' \emph{Oceanography}, vol.~26, no.~2, pp. 20--33, 2013. [Online]. Available: \url{http://www.jstor.org/stable/24862033}
\BIBentrySTDinterwordspacing

\bibitem{region_smoothing}
\BIBentryALTinterwordspacing
R.~Shang, J.~Lin, L.~Jiao, and Y.~Li, ``Sar image segmentation using region smoothing and label correction,'' \emph{Remote. Sens.}, vol.~12, p. 803, 2020. [Online]. Available: \url{https://api.semanticscholar.org/CorpusID:215414221}
\BIBentrySTDinterwordspacing

\bibitem{region_growing}
M.~Merzougui and A.~El~allaoui, ``Region growing segmentation optimized by evolutionary approach and maximum entropy,'' \emph{Procedia Computer Science}, vol. 151, pp. 1046--1051, 01 2019.

\bibitem{clustering}
D.~Nameirakpam, K.~Singh, and Y.~Chanu, ``Image segmentation using k -means clustering algorithm and subtractive clustering algorithm,'' \emph{Procedia Computer Science}, vol.~54, pp. 764--771, 12 2015.

\bibitem{fuzzy_clustering}
X.~Jia, T.~Lei, X.~Du, S.~Liu, H.~Meng, and A.~K. Nandi, ``Robust self-sparse fuzzy clustering for image segmentation,'' \emph{IEEE Access}, vol.~8, pp. 146\,182--146\,195, 2020.

\bibitem{watershed}
I.~Levner and H.~Zhang, ``Classification-driven watershed segmentation,'' \emph{IEEE Transactions on Image Processing}, vol.~16, no.~5, pp. 1437--1445, 2007.

\bibitem{graph_based}
P.~Felzenszwalb and D.~Huttenlocher, ``Efficient graph-based image segmentation,'' \emph{International Journal of Computer Vision}, vol.~59, pp. 167--181, 09 2004.

\bibitem{los_alamos}
B.~Tasseff, T.~Albash, Z.~Morrell, M.~Vuffray, A.~Y. Lokhov, S.~Misra, and C.~Coffrin, ``On the emerging potential of quantum annealing hardware for combinatorial optimization,'' 2022.

\bibitem{ATR_SAR}
O.~Kechagias-Stamatis and N.~Aouf, ``Automatic target recognition on synthetic aperture radar imagery: A survey,'' \emph{IEEE Aerospace and Electronic Systems Magazine}, vol.~36, no.~3, pp. 56--81, 2021.

\bibitem{graph_coloring}
M.~Demange, T.~Ekim, B.~Ries, and C.~Tanasescu, ``On some applications of the selective graph coloring problem,'' \emph{European Journal of Operational Research}, vol. 240, pp. 307--314, 01 2015.

\bibitem{graph_cut}
C.~Arora, S.~Banerjee, P.~Kalra, and S.~Maheshwari, ``An efficient graph cut algorithm for computer vision problems,'' vol. 6313, 09 2010, pp. 552--565.

\bibitem{mrf_image_kato}
\BIBentryALTinterwordspacing
Z.~Kato, J.~Zerubia \emph{et~al.}, \emph{Markov random fields in image segmentation}.\hskip 1em plus 0.5em minus 0.4em\relax Now Publishers, Inc., 2012, vol.~5, no. 1--2. [Online]. Available: \url{https://inria.hal.science/hal-00737058}
\BIBentrySTDinterwordspacing

\bibitem{algo_EM}
F.~Dellaert, ``The expectation maximization algorithm,'' \emph{College of Computing, Georgia Institute of Technology}, 2002.

\bibitem{bayes}
M.~J. Zyphur and F.~L. Oswald, ``Bayesian estimation and inference: A user's guide,'' \emph{Journal of Management}, vol.~41, no.~2, pp. 390--420, Feb. 2015.

\bibitem{MRF_cliques}
Z.~Wu, D.~Lin, and X.~Tang, ``Deep markov random field for image modeling,'' \emph{Computer Vision -- ECCV 2016}, vol. 9912, pp. 295--312, 10 2016.

\bibitem{MRF_nphard}
S.~Geng, Z.~Kuang, J.~Liu, S.~Wright, and D.~Page, ``Stochastic learning for sparse discrete markov random fields with controlled gradient approximation error,'' NIH Public Access, p. 156, 2018.

\bibitem{belief_propagation}
S.~Xu, J.-Q. Han, L.~Zhao, and G.-H. Liu, ``Efficient belief propagation for image segmentation based on an adaptive mrf model,'' \emph{2013 IEEE 11th International Conference on Dependable, Autonomic and Secure Computing}, pp. 324--329, 2013.

\bibitem{variational_method}
Y.~Tian and Y.~Xue, ``Variation method overview of image segmentation,'' \emph{Journal of Physics: Conference Series}, vol. 1487, p. 012012, 03 2020.

\bibitem{potts}
X.~Wang and J.~Zhao, ``Image segmentation using improved potts model,'' in \emph{2008 Fourth International Conference on Natural Computation}, vol.~7, 2008, pp. 352--356.

\bibitem{cauchy}
T.~Wan, N.~Canagarajah, and A.~Achim, ``Segmentation of noisy colour images using cauchy distribution in the complex wavelet domain,'' \emph{Image Processing, IET}, vol.~5, pp. 159 -- 170, 04 2011.

\bibitem{huber}
O.~Gutiérrez, J.~De~la Rosa, J.~d.~J. Hernández, E.~González, and N.~Escalante, ``Semi-huber potential function for image segmentation,'' \emph{Optics express}, vol.~20, pp. 6542--54, 03 2012.

\bibitem{convergence_EM}
\BIBentryALTinterwordspacing
C.~Wu, C.~Yang, H.~Zhao, and J.~Zhu, ``On the convergence of the em algorithm: A data-adaptive analysis,'' 2016. [Online]. Available: \url{https://api.semanticscholar.org/CorpusID:55245903}
\BIBentrySTDinterwordspacing

\bibitem{init_EM}
E.~Shireman, D.~Steinley, and M.~Brusco, ``Examining the effect of initialization strategies on the performance of gaussian mixture modeling,'' \emph{Behavior Research Methods}, vol.~49, 12 2015.

\bibitem{usup_adaptative}
Z.~Kato, J.~Zerubia, M.~Berthod, and W.~Pieczynski, ``Unsupervised adaptive image segmentation,'' in \emph{1995 International Conference on Acoustics, Speech, and Signal Processing}, vol.~4, 1995, pp. 2399--2402 vol.4.

\bibitem{multilayer_mrf}
Z.~Kato, T.-C. Pong, and S.~Qiang, ``Unsupervised segmentation of color textured images using a multilayer mrf model,'' in \emph{Proceedings 2003 International Conference on Image Processing (Cat. No.03CH37429)}, vol.~1, 2003, pp. I--961.

\bibitem{grabcut}
Y.~Zhang, Y.~Jiazheng, L.~Hongzhe, and L.~Qing, ``Grabcut image segmentation algorithm based on structure tensor,'' \emph{The Journal of China Universities of Posts and Telecommunications}, vol.~24, pp. 38--47, 04 2017.

\bibitem{alpha_expansion}
M.~Schmidt and K.~Alahari, ``Generalized fast approximate energy minimization via graph cuts: Alpha-expansion beta-shrink moves,'' 2011.

\bibitem{k_means}
\BIBentryALTinterwordspacing
T.~Zhang, R.~Ramakrishnan, and M.~Livny, ``Birch: An efficient data clustering method for very large databases,'' \emph{SIGMOD Rec.}, vol.~25, no.~2, p. 103–114, jun 1996. [Online]. Available: \url{https://doi.org/10.1145/235968.233324}
\BIBentrySTDinterwordspacing

\bibitem{gauss_explicit}
\BIBentryALTinterwordspacing
F.~Pascal, L.~Bombrun, J.-Y. Tourneret, and Y.~Berthoumieu, ``Parameter estimation for multivariate generalized gaussian distributions,'' \emph{{IEEE} Transactions on Signal Processing}, vol.~61, no.~23, pp. 5960--5971, dec 2013. [Online]. Available: \url{https://doi.org/10.1109%2Ftsp.2013.2282909}
\BIBentrySTDinterwordspacing

\bibitem{weib_implicit}
A.~Al-Wakeel, A.~Razali, and A.~Mahdi, ``Estimation accuracy of weibull distribution parameters,'' \emph{Journal of Applied Sciences Research}, vol.~5, p. 790, 07 2009.

\bibitem{SAR_weibull}
G.~Gao, ``Gao, g.: Statistical modeling of sar images: A survey. sensors 10, 775-795,'' \emph{Sensors}, vol.~10, 01 2010.

\bibitem{SAR_weibull_2}
A.~Jain and D.~Singh, ``An optimal selection of probability distribution functions for unsupervised land cover classification of palsar-2 data,'' \emph{Advances in Space Research}, vol.~63, 09 2018.

\bibitem{D-Wave}
M.~Johnson, M.~Amin, S.~Gildert, T.~Lanting, F.~Hamze, N.~Dickson, R.~Harris, A.~Berkley, J.~Johansson, P.~Bunyk, E.~Chapple, C.~Enderud, J.~Hilton, K.~Karimi, E.~Ladizinsky, N.~Ladizinsky, T.~Oh, I.~Perminov, C.~Rich, and G.~Rose, ``Quantum annealing with manufactured spins,'' \emph{Nature}, vol. 473, pp. 194--8, 05 2011.

\bibitem{pairwise_ponderation}
\BIBentryALTinterwordspacing
K.~Petersen, J.~Fehr, and H.~Burkhardt, ``Fast generalized belief propagation for map estimation on 2d and 3d grid-like markov random fields,'' p. 41–50, 2008. [Online]. Available: \url{https://doi.org/10.1007/978-3-540-69321-5_5}
\BIBentrySTDinterwordspacing

\bibitem{temps_annealing}
O.~Galindo and V.~Kreinovich, ``What is the optimal annealing schedule in quantum annealing,'' pp. 963--967, 12 2020.

\bibitem{em_hard_1}
Y.~Zhang, M.~Brady, and S.~Smith, ``Segmentation of brain mr images through a hidden markov random field model and the expectation-maximization algorithm,'' \emph{IEEE Transactions on Medical Imaging}, vol.~20, no.~1, pp. 45--57, 2001.

\bibitem{em_hard_2}
Z.~Kato and T.-C. Pong, ``A markov random field image segmentation model for color textured images,'' \emph{Image and Vision Computing}, vol.~24, pp. 1103--1114, 10 2006.

\bibitem{MSTAR}
``Mstar dataset,'' \url{https://www.sdms.afrl.af.mil/index.php?collection=mstar}, accessed: 2023-05-02.

\bibitem{SSDD_dataset}
T.~Zhang, X.~Zhang, J.~Li, X.~Xu, B.~Wang, X.~Zhan, Y.~Xu, X.~Ke, T.~Zeng, H.~Su, I.~Ahmad, D.~Pan, C.~Liu, Y.~Zhou, S.~JUN, and S.~Wei, ``Sar ship detection dataset (ssdd): Official release and comprehensive data analysis,'' \emph{Remote Sensing}, vol.~13, p. 3690, 09 2021.

\bibitem{reflexion_annealing}
\BIBentryALTinterwordspacing
S.~Yarkoni, E.~Raponi, T.~Bäck, and S.~Schmitt, ``Quantum annealing for industry applications: introduction and review,'' \emph{Reports on Progress in Physics}, vol.~85, no.~10, p. 104001, sep 2022. [Online]. Available: \url{https://doi.org/10.1088%2F1361-6633%2Fac8c54}
\BIBentrySTDinterwordspacing

\bibitem{pasqal}
\BIBentryALTinterwordspacing
``Pasqal quantum computer.'' [Online]. Available: \url{https://www.pasqal.com/}
\BIBentrySTDinterwordspacing

\end{thebibliography}

\end{document}